\documentclass{aa}
\usepackage{graphics,epsfig,lscape}

\usepackage{natbib}
\bibpunct{(}{)}{;}{a}{}{,}

\newcommand{\ergcm}[1]{$\cdot 10^{#1}$ erg cm$^{-2}$ s$^{-1}$}

\newcommand{\hcm}[1]{$\cdot 10^{#1}$ cm$^{-2}$}
\newcommand{\ohcm}[1]{$10^{#1}$ cm$^{-2}$}
\newcommand{\expo}[1]{$\cdot 10^{#1}$}
\newcommand{\nh}{n_{\rm H}}

\newcommand{\ct}{cts s$^{-1}$}
\newcommand{\mv}{\hbox{$m_{\rm V}$}}
\newcommand{\mr}{\hbox{$m_{\rm R}$}}
\newcommand{\mb}{\hbox{$m_{\rm B}$}}
\newcommand{\rxj}{\hbox{\object{RX\,J0806.4-4123}}}
\begin{document}
 
\title{XMM-Newton observations of the isolated neutron star \rxj
        \thanks{XMM-Newton is an ESA Science Mission
               with instruments and contributions directly funded by ESA Member
               states and the USA (NASA)}
}
 
\author{F.~Haberl \and V.E.~Zavlin}

\titlerunning{An XMM-Newton observation of the isolated neutron star \rxj}
\authorrunning{Haberl \& Zavlin}
 
\offprints{F. Haberl, \email{fwh@mpe.mpg.de}}
 
\institute{Max-Planck-Institut f\"ur extraterrestrische Physik,
           Giessenbachstra{\ss}e, 85748 Garching, Germany}
 
\date{Received date; accepted date}
 
\abstract{
The isolated neutron star \rxj\ was observed with XMM-Newton in November 2000.
The data from the three EPIC instruments allowed us 
(i) to derive an improved X-ray position to an accuracy of 2--3\arcsec, (ii) to 
accumulate the first medium-resolution soft X-ray spectra of high statistical 
quality and (iii) to find a candidate for the neutron star rotation period. 
Although this period of 11.3714 s is formally detected at a 
3.5$\sigma$ level in 
the EPIC-pn data, the similar pulse profiles deduced from all three EPIC instruments 
increase the confidence that the period 
is real. The pulsed fraction of $\sim$6\% would then be the weakest 
X-ray flux modulation detected from dim isolated neutron stars. We fitted
the X-ray spectra with blackbody and neutron star atmosphere models and discuss
the results with respect to the brightness limit placed by optical images.
The reduced size of the error circle on the X-ray position should allow
deeper searches for an optical counterpart.
}
 
\maketitle
 
\section{Introduction}

The ROSAT all-sky survey source \rxj\ (\object{1RXS\,J080623.0-412233}) 
was suggested as candidate for an isolated neutron star (INS)
by \citet{1998AN....319...97H} based on X-ray properties of this source 
similar to the established cases \hbox{\object{RX\,J1856.5-3754}} 
\citep{1996Natur.379..233W} 
and \hbox{\object{RX\,J0720.4-3125}} \citep{1997A&A...326..662H}.
While the latter two (the brightest in the sample of 
INSs discovered with ROSAT) were optically identified with \mv\ $\sim25.6$ 
\citep{1997Natur.389..358W} and \mb\ $\sim26$ 
\citep{1998A&A...333L..59M,1998ApJ...507L..49K} blue objects, respectively, optical 
observations of \rxj\ were only able to derive a limiting B magnitude 
\mb\ $>24$ for a possible optical counterpart \citep{1998AN....319...97H}. 
High resolution X-ray spectra obtained 
from RX\,J1856.5-3754 and RX\,J0720.4-3125 have revealed
no evidence for photospheric absorption or emission features
and are well modeled as blackbody radiation with little interstellar absorption
\citep{2001A&A...365L.298P,2001A&A...379L..35B}.

Today seven ROSAT discovered INSs (hereafter dim INS) 
are known \citep[see references in][]{1999AIP...599..244M,2001A&A...378L...5Z}
with RX\,J0720.4-3125 
\citep[rotation period $P$=8.39 s;][]{1997A&A...326..662H},
RX\,J0420.0-5022 \citep[$P$=22.7 s;][]{1999A&A...351L..53H}
and  RBS 1223 \citep[$P$=5.16 s;][]{2002A&A...381...98H} 
showing pulsations in the X-ray flux. 
No radio emission has been detected yet from these objects.
While little doubt is left that these objects are indeed INSs, 
the origin of their X-rays is still a matter of debate. If we 
see thermal emission from young cooling neutron stars,
the observed blackbody temperatures in the range of 50--120 eV
imply ages younger than $\sim10^6$ 
years for standard cooling curves. Older neutron stars may be re-heated by accretion of
interstellar matter, producing X-ray luminosities sufficiently high to be observable.
However, the accretion rate strongly depends on the velocity of the neutron star relative
to the interstellar medium, and the neutron star must rotate slowly enough 
to allow the matter to enter the magnetosphere 
\citep[see the review of][]{2000PASP..112..297T}.
The recent estimates of a pulse period derivative of 
$(0.7-2.0)\cdot10^{-11}$ s s$^{-1}$ from RBS\,1223 \citep{2002A&A...381...98H}
and $\sim10^{-14}$ s s$^{-1}$ from RX\,J0720.4-3125 \citep{2002MNRAS.........Z}
with inferred magnetic field strengths of 
$\sim$2\expo{14}~G and $\sim10^{13}$ G, respectively, favour 
the models invoking a young (or middle-aged)
cooling neutron star rather than
the accretion heating, at least for these pulsating sources\footnote{
\citet{Kaplan} derived a less stringent upper limit for $\dot{P}$ for
RX\,J0720.4-3125 by not including XMM-Newton data, 
but also their results support the interpretation that this source
is a cooling neutron star.}.
By virtue of this, the dim nearby neutron
stars are suggested to be different from anomolous
X-ray pulsars which have been proposed to be powered
either by superstrong magnetic fields \citep{1996ApJ...473..322T}
or accretion \citep{1995ApJ...442L..17M}. 

and are well modeled as blackbody radiation with little interstellar absorpt
Following the spin history of the three known pulsars among the seven dim
INSs and searching for pulsations in the others with new sensitive instruments is
important for the understanding of the origin of their X-ray emission. Here we report
on an XMM-Newton observation of the poorly known INS \rxj. We present an improved X-ray
position, the X-ray spectrum and the detection of a 
candidate for the rotation period of the star.

\section{The XMM-Newton observation}

XMM-Newton \citep{2001A&A...365L...1J} observed \rxj\ during the
satellite revolution \#168 on 
November 8, 2000. Here we report on the analysis of the data 
collected with the European Photon Imaging Cameras (EPICs)
based on MOS \citep[EPIC-MOS1 and -MOS2,][]{2001A&A...365L..27T}
and pn \citep[EPIC-pn,][]{2001A&A...365L..18S} 
CCD detectors which are mounted behind the three X-ray telescopes 
\citep{2000SPIE.4012..731A}. All cameras were
operated in the Full-Frame mode (FF) providing data over a field of view of
$\sim$13\arcmin\ radius. Further details of the EPIC observations are summarized in 
Table~\ref{epic-obs}. The data were processed using the XMM-Newton analysis package 
SAS version 5.2 to produce the photon event files and tools from version 5.3
for further spatial, spectral and timing analysis. The event times were corrected to
solar barycenter using the SAS task ``barycen 1.13.1'' (in which a timing
problem present in older versions has been solved).

\begin{table}
\caption[]{XMM-Newton EPIC observations of \rxj\ on November 8, 2000.}
\begin{tabular}{clcccc}
\hline\noalign{\smallskip}
\multicolumn{1}{l}{Camera} &
\multicolumn{1}{c}{Read-out} &
\multicolumn{1}{c}{Filter} &
\multicolumn{2}{c}{Observation} &
\multicolumn{1}{c}{Exp.} \\
 &
\multicolumn{1}{c}{Mode} &
 &
\multicolumn{1}{c}{Start} &
\multicolumn{1}{c}{End (UT)} &
\multicolumn{1}{c}{[ks]} \\

\noalign{\smallskip}\hline\noalign{\smallskip}
 MOS1/2  & FF, 2.6 s & Thin & 13:43 & 18:47 & 18.0 \\
 pn      & FF, 73 ms & Thin & 14:24 & 18:53 & 15.6 \\
\noalign{\smallskip}
\hline
\end{tabular}
\label{epic-obs}
\end{table}

\subsection{The X-ray position of \rxj}

\begin{figure}
\resizebox{\hsize}{!}{\includegraphics[clip=]{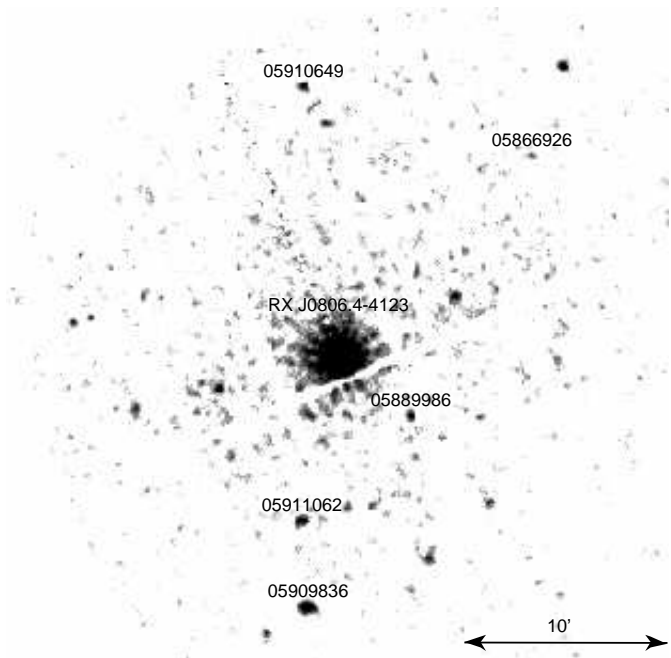}}
\caption{EPIC-pn image in the 0.2--2.0 keV band. The image is smoothed with an
         adaptive intensity filter and out-of-time events (caused by photons
         detected during CCD-frame read-out) are subtracted on a statistical basis. 
         Marked are the identified USNO A2.0 objects (see Table~\ref{epic-usno})
         used to define the coordinate reference frame.}
\label{epic-ima}
\end{figure}

\citet{1998AN....319...97H} derived an error radius of 11\arcsec\ (90\% confidence) 
from the ROSAT PSPC observations of \rxj. The error is dominated by systematic 
uncertainties in the attitude reconstruction which could not be reduced due to the 
lack of other X-ray sources with optical counterparts detected in the field of view 
to define a reference coordinate frame.
In order to find field stars in the EPIC images, a standard source detection analysis using 
combined box sliding and maximum likelihood techniques as available in SAS, was performed
simultaneously in the three energy bands 
0.3--1.0, 1.0--2.0 and 2.0--7.5 keV.
This yielded 50, 29 and 33 detections with existence likelihood larger than 10 in the pn, MOS1 
and MOS2 images, respectively. Fig.~\ref{epic-ima} shows an EPIC-pn soft X-ray
image. To identify 
possible optical counterparts to the X-ray sources, finding charts were produced using the 
DSS2 (red) image. To secure the identification, the following criteria were applied:
(i) only X-ray sources which were detected by at least two EPIC instruments were used, 
(ii) a stellar-like object 
was found in the X-ray error circle, 
and (iii) both X-ray (spectral) and optical properties (R and B 
magnitudes as derived from the USNO A2.0 catalogue produced by the US Naval Observatory) 
must be compatible with a star. This resulted in the identification of five X-ray sources 
with field stars as summarized in Table~\ref{epic-usno}. 
Hardness ratios are defined as 
HR1 = (B+A)/(B-A) and HR2 = (C+B)/(C-B) with A, B and C as count rates in the energy bands
0.3--1.0, 1.0--2.0 and 2.0--7.5 keV, respectively. Coronal emission from stars is 
characterized by relatively soft X-ray spectra with 
the bulk of emission below 2 keV. This results in negative values for both hardness 
ratios with HR1 depending on the amount of interstellar
absorption \citep[see also][]{2002ESASP488H}. HR2 values of weak sources are often 
undetermined due to the lack of statistics in the 2.0--7.5 keV band. For comparison 
the X-ray properties of the target are also shown in Table~\ref{epic-usno}.

\begin{table*}
\caption[]{Field stars near \rxj\ observed with EPIC pn.}
\begin{tabular}{ccccccc}
\hline\noalign{\smallskip}
\multicolumn{1}{l}{ML$^1$} &
\multicolumn{1}{c}{count rate$^2$} &
\multicolumn{1}{c}{HR1} &
\multicolumn{1}{c}{HR2} &
\multicolumn{1}{c}{USNO counterpart} &
\multicolumn{1}{c}{\mr} &
\multicolumn{1}{c}{\mb} \\
& 
\multicolumn{1}{c}{[10$^{-3}$ s$^{-1}$}] &
& & &
\multicolumn{1}{c}{[mag]} &
\multicolumn{1}{c}{[mag]} \\
\noalign{\smallskip}\hline\noalign{\smallskip}
 518         & 30.0$\pm$2.5 & -0.37$\pm$0.08 & -0.68$\pm$0.15 & U0450\_05909836 & 13.0 & 14.0 \\
 124         &  8.5$\pm$1.1 & -0.22$\pm$0.13 & --             & U0450\_05911062 & 15.0 & 16.0 \\
 74          &  8.4$\pm$1.4 & -0.41$\pm$0.15 & -0.11$\pm$0.32 & U0450\_05910649 & 15.3 & 16.5 \\
 46          &  3.7$\pm$0.8 & -0.61$\pm$0.19 & --             & U0450\_05889986 & 17.3 & 19.2 \\
 28          &  4.8$\pm$1.2 & -0.17$\pm$0.26 & --             & U0450\_05866926 & 13.3 & 14.5 \\
\noalign{\smallskip}\hline\noalign{\smallskip}
 $10^5$ &  1060$\pm$10 & -0.96$\pm$0.01 & -0.92$\pm$0.06 & --	        &  --  &  --  \\
\noalign{\smallskip}
\hline
\end{tabular}

$^1$ Likelihood for existence with probability $p_{\rm ML} = 1-\exp(-{\rm ML})$\\
$^2$ 0.3--2.0 keV
\label{epic-usno}
\end{table*}

\begin{figure}
\resizebox{\hsize}{!}{\includegraphics[clip=]{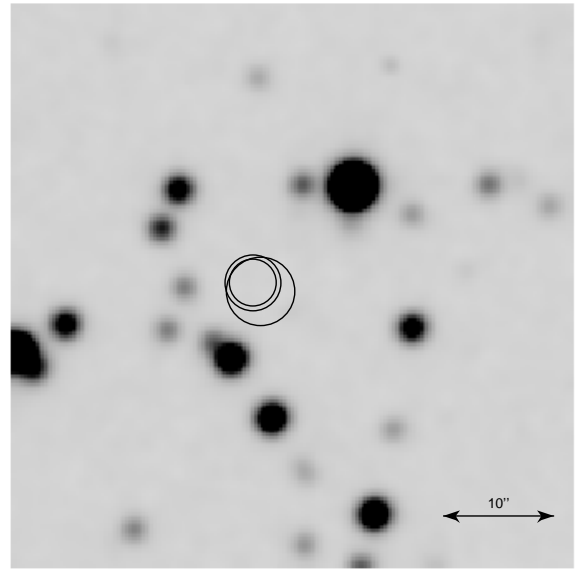}}
\caption{B-band CCD image around the X-ray position of \rxj\ 
         \citep[from][]{1998AN....319...97H}. The circles represent the position
	 uncertainties derived from pn (radius 2.1\arcsec), MOS1 (3.1\arcsec) and MOS2
	 (2.6\arcsec) which include statistical and (dominating) systematic X-ray errors 
	 and a 0.25\arcsec\ error from the USNO A2.0 positions.}
\label{epic-eso}
\end{figure}

Using the SAS task ``eposcorr'' the X-ray source positions were corrected to the USNO A2.0 
catalogue reference frame by minimizing the differences in X-ray to optical position 
for the five field stars. The uncertainties on the required coordinate shifts can be regarded 
as the remaining systematic error of the X-ray bore-sight and are 2.1\arcsec, 3.1\arcsec\ and 
2.5\arcsec\ for pn, MOS1 and MOS2, respectively. Fig.~\ref{epic-eso} shows the B-band image of
the area around \rxj\ as obtained by \citet{1998AN....319...97H} with the three error circles 
derived from the EPIC instruments after the bore-sight correction. The corresponding position
deduced from the pn image is RA = 08$^{\rm h}$06$^{\rm m}$23\fs47 and 
Dec = --41\degr22\arcmin32\farcs3 (J2000.0). The error circles obtained from the
EPIC data agree with each other and are completely inside the ROSAT circle. This 
confirms that none of the surrounding stars is responsible for the X-ray emission 
from \rxj. 

\subsection{The temporal behaviour of \rxj}

\begin{figure}
\resizebox{\hsize}{!}{\includegraphics[angle=-90,clip=]{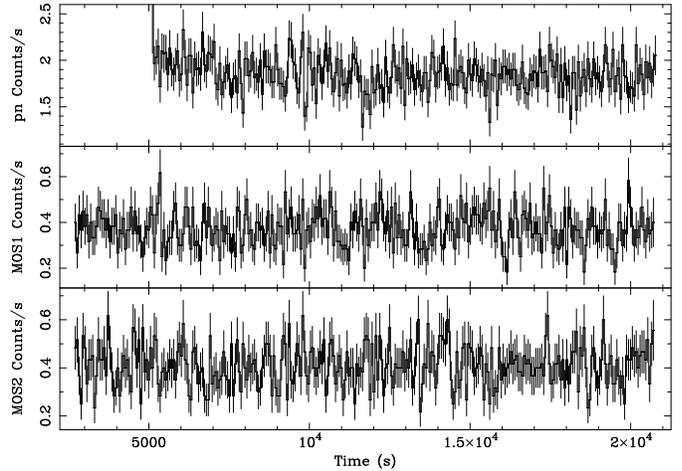}}
\caption{EPIC light curves with a time resolution of 60 s derived from the 0.12--1.2 keV 
         energy band. The background with average values of 3.9$\cdot10^{-2}$ \ct, 
	 3.1$\cdot10^{-3}$ \ct\ and 4.5$\cdot10^{-3}$ \ct\ for pn, MOS1 and MOS2, 
	 respectively, is not subtracted. Start time is 2000, Nov. 8, 13:45:12 UT.}
\label{epic-lcurve}
\end{figure}

\begin{figure}
\resizebox{\hsize}{!}{\includegraphics[angle=-0,clip=]{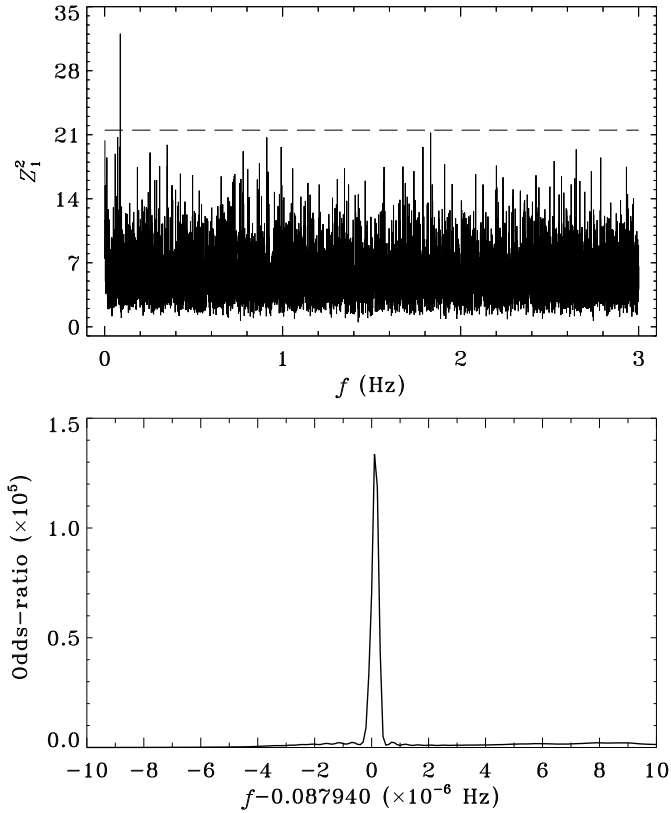}}
\caption{Power spectrum (top panel) and frequency
dependence of the Odds-ratio. The dashed line in the top panel 
shows a $Z^2_1=21.5$ level below which all peaks can be 
regarded as noise with a probability of 100\%.
}
\label{power-odds}
\end{figure}

EPIC light curves of \rxj\ were extracted in the 0.12--1.2 keV energy band (using events 
with pixel patterns 0-12) which covers most of the X-ray spectrum (see 
the next Section). 
The light curves show no significant variations over the duration of the observation with 
average net count rates of 1.81 \ct, 0.369 \ct\ and 0.405 \ct\ in the
pn, MOS1 and MOS2 detectors, respectively (Fig.~\ref{epic-lcurve}). 
To search for periodicities on short time scales,
we used 29,549 events extracted from a circle with radius of 45\arcsec\
centered on the position of the target in the pn data. 
Of these selected events, about 97.5\% were estimated to belong 
to the source. As a periodicity test we used the Rayleigh
$Z^2_n$ method \citep{1983A&A...128..245B} 
with one harmonic involved ($n=1$) which is expected
to be optimal to search for rather smooth pulsations.
The 0.073~s time resolution of the pn data limits the 
periodicity search to frequencies below $\sim$ 3 Hz.
We performed the $Z_1^2$ test 
in the $\Delta f=0.001-3$ Hz frequency range with a step of
$5\cdot 10^{-6}$ Hz. The oversampling by a factor of 10
(compared to the expected widths of $1/T_{\rm sp}$ of the $Z^2_1$
peaks, where $T_{\rm sp}=15.7$ ks is the observational time span)
was chosen to resolve separate $Z^2_1$ peaks and to assure 
that no peak corresponding to a periodic signal is missing.
As the variable $Z^2_n$ has a probability density function
equal to that of $\chi^2$ with $2\cdot n$ degrees of freedom
\citep[e.~g.,][]{1986Wiley.B}, 
the probability to obtain a noise peak of a given $Z^2_1$ 
height in one trial
is $\exp(-Z^2_1/2)$. The number of independent trials
in the chosen frequency range estimated
as $N=\Delta f\, T_{\rm sp}=47.1\cdot 10^3$
puts a lower limit on a $Z^2_1$ value to be yielded
by a periodic signal with a probability $p>0$,
$Z^2_1 > 2\,\log(N)=21.5$ (peaks with $Z^2_1<21.5$ 
are regarded as noise with a probability of 100\%).
This test resulted in only one peak higher than that lower limit, 
$Z^2_1=33.9$ at the frequency $f_*=0.087936$ Hz
(see Fig.~\ref{power-odds}).
The probability to obtain by chance a peak of $Z^2_1=33.9$
in $N$ independent trials is 
$p=N\,\exp(-Z^2_1/2)=2.0\cdot 10^{-3}$,
which corresponds to a detection of pulsations at a confidence level
of $C=(1-p)\cdot 100\%=99.8\%$, or $3.1\sigma$.
Involving higher harmonics gives values 
$Z^2_2=42.9$, $Z^2_3=43.0$
and $Z^2_4=43.2$ at the same frequency $f_*$. 
These numbers indicate that, according to the $H$-test suggested by 
\citet{1989A&A...221..180D}, 
there is a statistically significant contribution 
from the second harmonic. A probability
to obtain a noise peak of $Z^2_2=42.9$ in one trial is only
$p=1.1\cdot 10^{-8}$ and in $N$ independent trials is 
$p=5.1\cdot 10^{-4}$, that increases the significance 
of the period detection up to a $3.5\sigma$ level.
We note this $Z^2_2$ peak is highest in the
frequency range chosen for the periodicity search
(all other $Z^2_2$ peaks are below 27.0 and produced
by noise with probabilities $p>94\%$).
We also performed the $Z^2_n$-test adding 14,210 source photons
detected with two MOS cameras to the pn data.
The increase of the total statistics by about 50\%
results only in small increments in $Z^2_n$ values
($Z^2_1=37.5$ and $Z^2_2=43.9$ at the same frequency $f_*$)
and significance of the detected signal (to a 3.6$\sigma$ level).
That can be explaind by the low time resolution of the MOS data
(see Table~\ref{epic-obs}) which should lead to a strong smearing 
of the weak signal.

To determine the pulsation frequency and its uncertainties
more accurately, we applied the Odds-ratio method by 
\citet{1996ApJ...473.1059G} 
based on the Bayesian formalism. The 
frequency-dependent Odds-ratio $O(f)$ specifies how the data
favour a periodic model of a frequency $f$ over the unpulsed
model. This ratio allows one to obtain the probability distribution function
$p(f)\propto O(f)f^{-1}$ for a signal to be periodic in a chosen 
frequency range \citep[see][ for more details]{2000ApJ...540L..25Z}. 
Fig.~\ref{power-odds} shows the dependence of $O(f)$ in a range around
the frequency $f_*$ resulting from the $Z^2_1$ test.
The maximum value 
$O_{\rm max}=1.3\cdot 10^5$
is at $f=f_0=0.087940$ Hz
($f_0-f_*=4\cdot 10^{-7}~{\rm Hz}\,\ll\, 1/T_{\rm sp}$).
The uncertainties of $f_0$ at the 68\% and 90\% 
confidence levels are 
$\delta f=2.3\cdot 10^{-6}$ and $10.5\cdot 10^{-6}$ Hz,
respectively. The corresponding rotation period is
$P_0=11.37139$ s with uncertainties of 
$3.0\cdot 10^{-4}$ and $13.6\cdot 10^{-4}$ s
at the 68\% and 90\% levels, respectively.

\begin{figure}
\resizebox{\hsize}{!}{\includegraphics[angle=-0,clip=]{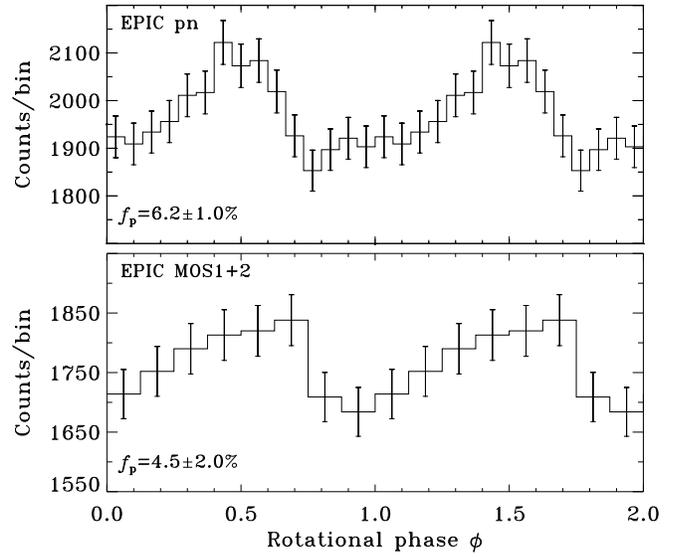}}
\caption{
  Pulse profiles extracted from the pn (top panel) and combined
  MOS1 and MOS2 data from \rxj\ at $f=f_0=0.087940$ Hz.} 
\label{epic-efold}
\end{figure}

The folded light curve extracted at $f=f_0$ from the pn data (see Fig.~\ref{epic-efold})
reveals one broad pulse per period with the pulsed fraction
of $f_p=6.2\pm 1.0\%$.
The
MOS light curve extracted 
at $f=f_0$ shows a remarkable similarity with that obtained from the
pn data (Fig.~\ref{epic-efold}),
although the MOS time resolution makes
its pulse somewhat broader and pulsed fraction  smaller
than those in the pn light curve.
We also note that the ROSAT PSPC observation of
\rxj\ in October 1993 yielded too scanty statistics 
(about 900 counts) for searching a signal of such a small
pulsed fraction (the signal should result in a too  
low peak with $Z^2_1\sim 1$, which cannot be discriminated from the noise).

\subsection{The X-ray spectrum of \rxj}

\begin{figure*}
\resizebox{9cm}{!}{\includegraphics[clip=]{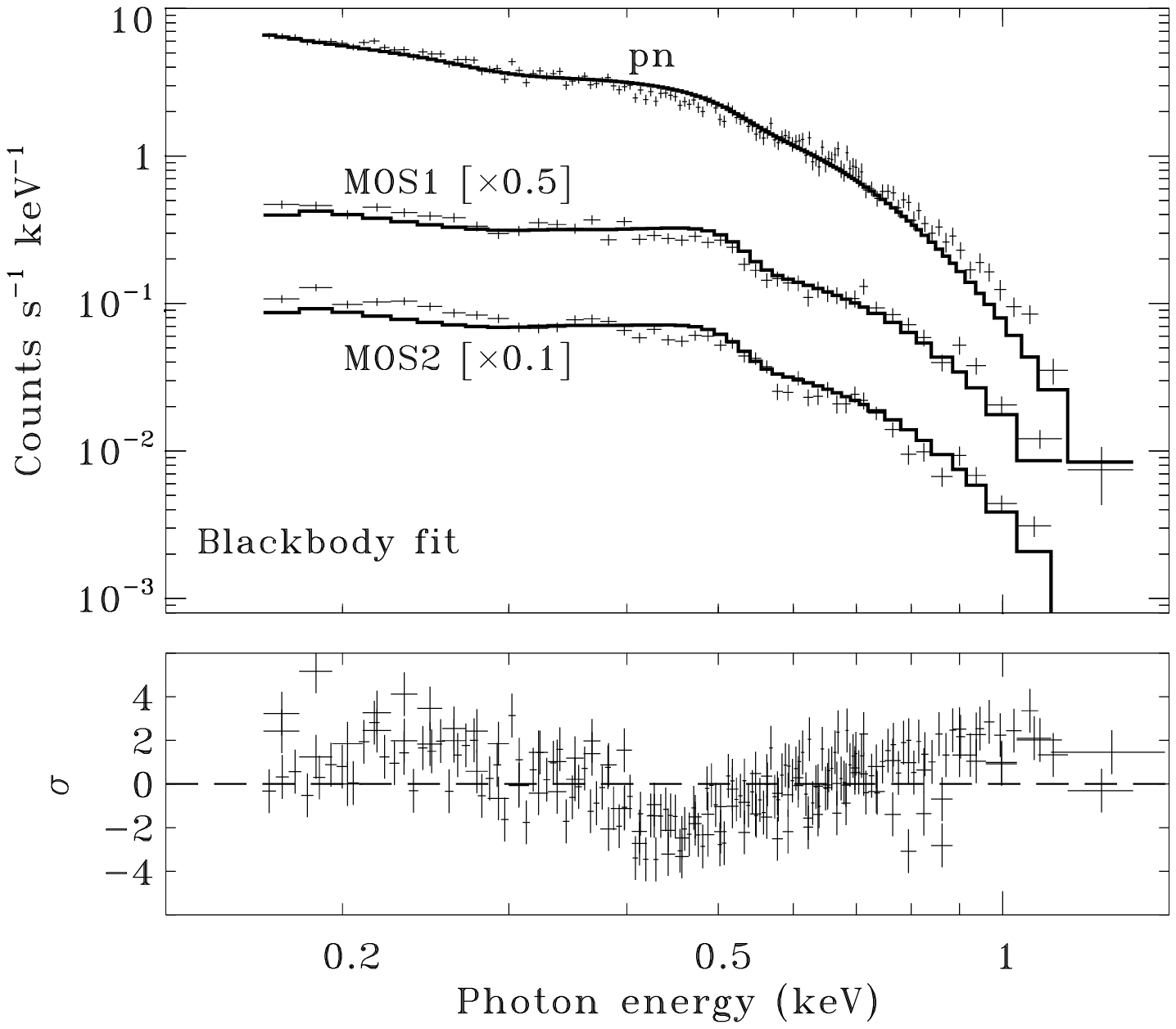}}
\resizebox{9cm}{!}{\includegraphics[clip=]{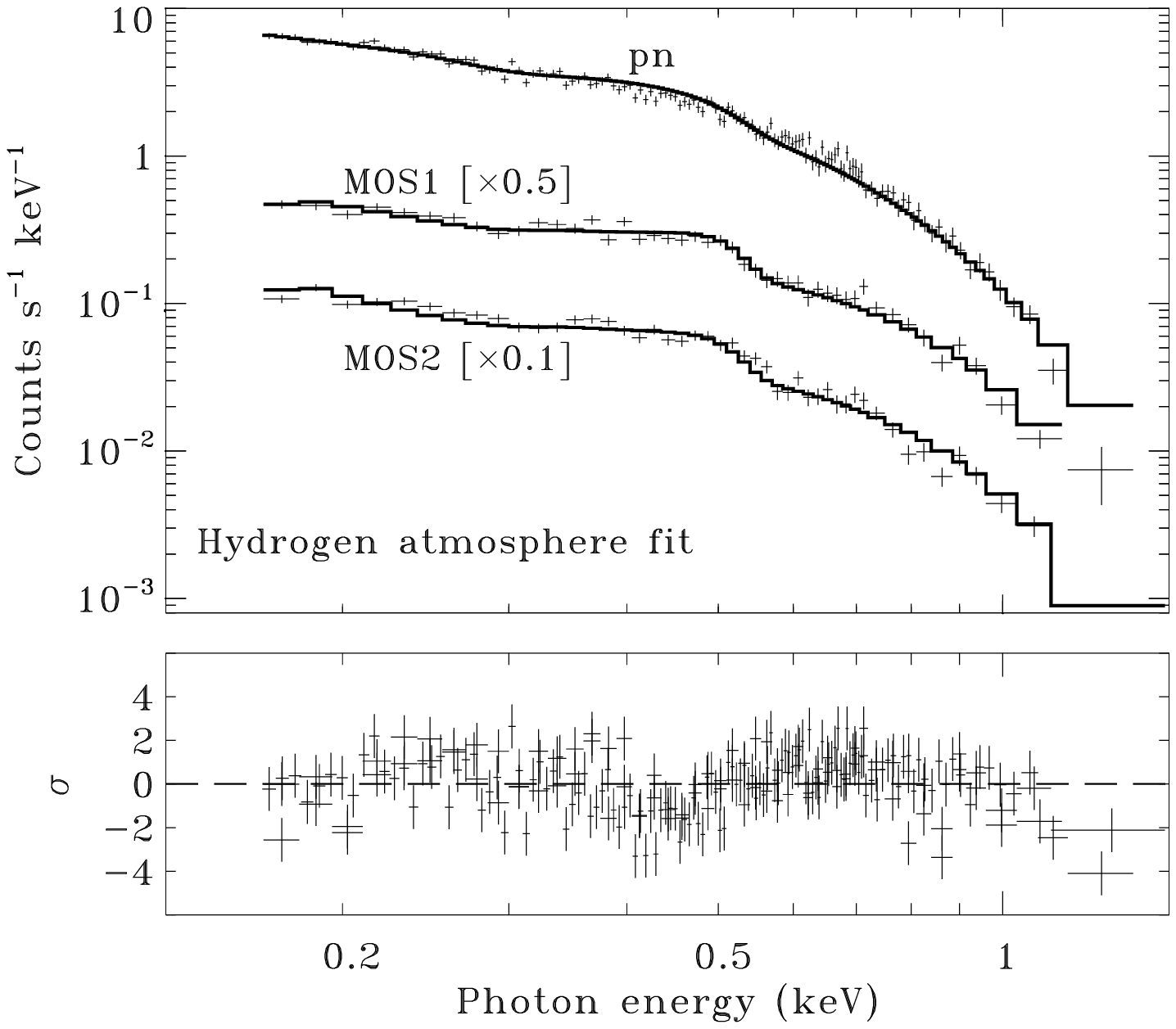}}
\caption{EPIC spectra of \rxj\ together with the best fit
         blackbody (left) and atmosphere (right) models. 
	 Most of the residuals (shown in the bottom panels) are probably caused by calibration
	 uncertainties around instrumental absorption edges.}
\label{epic-spectra}
\end{figure*}

The X-ray spectra were extracted using single- and double-pixel events 
(pattern 0-4) from pn and pattern 0-12 events from MOS data. 
Detector response matrix files were used as available from the
hardware groups in November 2001 (version 6.1 for pn and v9q20t5r6 for MOS).
First, an absorbed blackbody model with normalization parameters allowed to 
be free for the individual instruments was simultaneously fitted to the three 
spectra (left panels in Fig.~\ref{epic-spectra}).
From the best fit a blackbody temperature of 
$kT^\infty=94\pm$1 eV\footnote{
The superscript $^\infty$ is used to indicate that 
quantities are given as measured by a distant observer. Errors on the
spectral parameters are given at a 90\% confidence level.}
and an upper limit for the hydrogen column density
$\nh$ of \ohcm{18} were found
($\chi^2_\nu=2.9$ for 228 degrees of freedom, dof). The low value for the absorption 
may be caused by calibration uncertainties mainly below 0.5 keV. 
The radius of the emitting area derived from the
blackbody fit is $R^\infty\simeq 0.6\,(d/100\,{\rm pc})$~km.
Fitting the pn 
and MOS spectra separately yields blackbody temperatures of 
$kT^\infty=97\pm 1$ eV and $91\pm 1$ eV for the pn and 
MOS data, respectively ($\chi^2_\nu=2.5$ for 224 dof).
The model fluxes in the 0.1-2.4 keV range, 
2.82\ergcm{-12}, 3.08\ergcm{-12} and 3.38\ergcm{-12}
as derived from the combined fit for
the pn, MOS1 and MOS2 spectra (respectively),
show differences between the three instruments
larger than the typical statistical errors of 1--2\% 
obtained from the fit. For comparison, \citet{1998AN....319...97H} 
derived a flux of (2.9$^{+0.30}_{-0.15}$)\ergcm{-12} (90\% confidence) 
from the ROSAT PSPC data.

Nonmagnetic neutron star hydrogen atmosphere models 
\citep{1996A&A...315..141Z}\footnote{
The hydrogen atmosphere models are available at 
{\tt http://legacy.gsfc.nasa.gov/docs/xanadu/xspec}} 
fit the pn and MOS spectra fairly well, resulting
in the surface temperature $kT^\infty\simeq 26$ eV
and a 110 pc distance to \rxj\ assuming the ``canonical''
neutron star radius $R_{\rm NS}=10$ km and mass $M_{\rm NS}=1.4\,M_\odot$
($\chi^2_\nu=1.8$ for 228 dof; right panels in Fig.~\ref{epic-spectra}).
To account for systematic calibration uncertainties, $\nh$ was allowed 
to vary in fits to the spectra from the different 
instruments, while the surface temperature and distance were 
kept at the same values for all three spectra.
The obtained values of the hydrogen column density are
$\nh\simeq$ 1.6\hcm{20}, 1.0\hcm{20} and 0.4\hcm{20} for the 
fits to the pn, MOS1 and MOS2 data, respectively.
Neither pure iron nor solar-mixture atmosphere
models fit the EPIC spectra of \rxj\ ($\chi^2_\nu\simeq 20$).

\section{Discussion}

The XMM-Newton observations of the INS candidate \rxj\ revealed pulsations in its X-ray flux
with a period of 11.3714 s. This is well within the range of 5.16 s to 22.7 s observed from 
three other dim INSs. Although detected at a statistical level of $3.5\sigma$,
the similar pulse profiles seen in the data from  
all EPIC instruments inspire confidence that the modulation is
real and indicates the rotation period of the neutron star.
The pulsed fraction of 6.2\% is the weakest detected so far in X-rays from these 
objects (9\% in the X-ray flux of RX\,J0720.4-3125, \citet{2001A&A...365L.302C};
20$\pm$2\% in RBS\,1223, \citet{2002A&A...381...98H};
43$\pm$14\% in RX\,J0420.0-5022, \citet{1999A&A...351L..53H}). 
Such a modulation is too small to be found in available data of any of the other dim INSs. 
For example, only an upper limit of 4.5\% was 
derived from a 450 ks Chandra observation of RX\,J1856.5-3754
\citep{2002ApJ...570L..75R}. 
A simplest explanation for a small pulsed fraction in
X-ray radiation from an INS is that the rotational axis of the star
is nearly co-aligned with either the line of sight to a distant
observer or the magnetic axis of the star (or both).
Another reason resulting in the small pulsed fraction can be
the effect of light bending in the strong gravitational
field near the neutron star surface
\citep[e.~g.][]{1983ApJ...274..846P,1988ApJ...325..207R},
For instance, if a neutron star is rather massive, so that
$[M_{\rm NS}/M_\odot]/[R_{\rm NS}/1~{\rm km}] > 0.177$,
the whole star's surface becomes visible 
\citep[e.~g.][]{1995PAZh...21..168Z}, 
which should result in a strong smearing of the modulated X-ray
flux emitted by the neutron star (even if the radiation
originates from a small spot on the star's surface). However, assuming that 
radii and masses of the detected INSs are clustering around
the canonical values of 
$R_{\rm NS}=10$ km and $M_{\rm NS}=1.4\,M_\odot$,
we may conclude that 
the different observed pulsed fractions are mainly caused by geometrical
effects, namely, different orientations of the stars' axes and/or different
sizes of the emitting areas.

In this respect it is interesting that for the
two dim INSs with 
the estimates on
strength of the surface magnetic 
fields, $\sim$2\expo{13}~G for RX\,J0720.4-3125 
\citep{2002MNRAS.........Z} and 
$\sim$2\expo{14}~G for RBS\,1223 \citep{2002A&A...381...98H},
the pulsed fraction in the X-ray flux from the 
object with the stronger field is significantly higher, as is expected 
if the radiation is emitted from hot spots
around the magnetic poles. Hot polar caps can either be produced by 
a flux of relativistic particles falling down onto the polar caps 
from the pulsar magnetosphere or 
anisotropy of the thermal conductivity in the neutron star 
crust caused by a very strong magnetic field 
\citep[see, e.~g.,][]{1983ApJ...271..283G}. 
While it looks feasible to explain the shallow modulation observed from \rxj\ 
by the anisotropic heat conduction, \citet{2001A&A...365L.302C} 
argue that the relatively smooth temperature 
gradient on the neutron star surface produced 
by this mechanism alone can only marginally 
explain the pulse profile of RX\,J0720.4-3125. 
Higher pulsed fractions will be even more 
difficult to reconcile with this model,
unless the surface radiation is intrinsically anisotropic
as it may emerge from a magnetized neutron star atmosphere 
\citep{1994A&A...289..837P,1995A&A...297..441Z}. 
It is clear that whatever mechanism produces the hot polar regions,
measuring rotational period changes and, therefore, magnetic 
field strengths for the pulsating dim INS is crucial for understanding 
the origin of their X-ray emission.

The EPIC spectral data on \rxj\ are well consistent with a smooth,
featureless model and can be fitted with both blackbody
and neutron star hydrogen (or helium) atmosphere spectra
(since there are still fairly large uncertainties in the
calibration of the instrument responses at energies below 0.5-0.6 keV,
we do not consider the blackbody approach as rejected by
the data because of the rather poor quality of the spectral fit). 
The derived blackbody temperature of $\simeq$94 eV is within the 
range of values found for other X-ray dim neutron stars,
in particular, it is close to that of RX\,J1605.3+3249 
\citep{1999A&A...351..177M}. Extrapolating the best-fit blackbody model 
to the optical wavelength range gives a B magnitude
$\mb\simeq30.8$ (neglecting extinction)
for a possible optical counterpart -- much fainter flux than 
the upper limit put by \citet{1998AN....319...97H}.
It is worthwhile to mention that the 
optical counterparts found for the INSs
RX\,J1856.5-3754 and RX\,J0720.4-3125 
are significantly brighter than the extrapolations of the
blackbody spectra obtained from the X-ray data on
these two objects \citep{2002ApJ...564..981P,1998A&A...333L..59M}. 
The hydrogen atmosphere model fit predicts 
$\mb\simeq26.9$, still in agreement with the upper limit $\mb>24$.
On the other hand, optical fluxes predicted by the light-element 
atmosphere models for RX\,J1856.5-3754 
\citep{1996ApJ...472L..33P,2002ApJ...564..981P}
and RX\,J0720.4-3125 \citep{motch}
are much brighter than the actually measured values.
If \rxj\ has 
properties similar to those of these two INSs,
then the B magnitude of its optical flux is expected to be  
in the range of 28--30.
Utilizing the improved X-ray position derived from the XMM-Newton
data on \rxj, more stringent constraints on the 
neutron star parameters can be obtained from future deep optical observations.

\begin{acknowledgements}
The XMM-Newton project is supported by the Bundesministerium f\"ur Bildung und
For\-schung / Deutsches Zentrum f\"ur Luft- und Raumfahrt (BMBF / DLR), the
Max-Planck-Gesellschaft and the Heidenhain-Stif\-tung.
\end{acknowledgements}

\bibliographystyle{apj}
\bibliography{ins,general,myrefereed,myunrefereed,mytechnical}

\end{document}